\documentclass[journal]{IEEEtran}
\usepackage{amsmath,amsfonts}

\usepackage{algorithm}
\usepackage{algpseudocode}  
\usepackage{array}
\usepackage[caption=false,font=normalsize,labelfont=sf,textfont=sf]{subfig}
\usepackage{textcomp}
\usepackage{stfloats}
\usepackage{url}
\usepackage{dsfont}
\usepackage{verbatim}
\usepackage{graphicx}
\usepackage{cite}
\usepackage{tabularx} 
\usepackage{enumitem} 
\usepackage{times}  
\usepackage{xcolor} 
\usepackage{soul}
\setlist{nosep}

\usepackage[normalem]{ulem}

\begin{document}

\title{Mitigating Backdoor Triggered and Targeted Data Poisoning Attacks in Voice Authentication Systems}

\author{Alireza Mohammadi, Keshav Sood, Dhananjay Thiruvady, Asef Nazari

\IEEEcompsocitemizethanks{\IEEEcompsocthanksitem~Alireza Mohammadi, Keshav Sood, Dhananjay Thiruvady, and Asef Nazari are with Deakin University, Geelong, Australia. Email: alireza.mohammadi9207@gmail.com; keshav.sood@deakin.edu.au; dhananjay.thiruvady@deakin.edu.au; asef.nazari@deakin.edu.au.}}

\markboth{}%
{Shell \MakeLowercase{\textit{et al.}}: A Sample Article Using IEEEtran.cls for IEEE Journals}

\maketitle
\IEEEpubidadjcol

\begin{abstract}

Voice authentication systems remain susceptible to two major threats: backdoor-triggered attacks and targeted data poisoning attacks. This dual vulnerability is critical because conventional solutions typically address each threat type separately, leaving systems exposed to adversaries who can exploit both attacks simultaneously. We propose a unified defense framework that effectively addresses both BTA and TDPA. Our framework integrates a frequency-focused detection mechanism that flags covert pitch-boosting and sound-masking backdoor attacks in near real-time, followed by a convolutional neural network that addresses TDPA. This dual-layered defense approach utilizes multidimensional acoustic features to isolate anomalous signals without requiring costly model retraining. In particular, our PBSM detection mechanism can seamlessly integrate into existing voice authentication pipelines and scale effectively for large-scale deployments. Experimental results on benchmark datasets and their compression with the state-of-the-art (SoTA) algorithm demonstrate that our PBSM detection mechanism outperforms the SoTA. 
Our framework reduces attack success rates to as low as 5–15\%, while maintaining a recall rate of up to 95\% in recognizing TDPA.

\end{abstract}


\section{Introduction}
\label{sec:intro}

\IEEEPARstart{V}{oice} authentication systems (VAS)—often referred to as speaker recognition or voice biometrics—identify or verify users by analyzing distinctive acoustic characteristics in their speech~\cite{10.1145/3701985}. These systems typically operate in two phases: \textit{enrollment}, where a user’s voice profile is recorded and stored, and \textit{verification}, where an incoming voice sample is compared against the stored profile. Due to their ease of use and hands-free operation, voice authentication systems have gained widespread adoption in many domains including mobile banking, call center authentication, digital assistants (e.g., Siri, Alexa), and enterprise security solutions. However, the reliability of such systems critically depends on their security. If adversaries successfully exploit underlying vulnerabilities, they can subvert legitimate user profiles or bypass verification, fundamentally undermining the trustworthiness of voice biometrics \cite{10463411,10179374}. \par

\footnote{Table \ref{tab:abbreviations} provides a summary of key abbreviations.}

\begin{table}[ht]
    \caption{List of Abbreviations}
    \label{tab:abbreviations}
    \centering
    \renewcommand{\arraystretch}{1.2}
    \begin{tabularx}{\columnwidth}{|c|X|}
        \hline
            \textbf{Abbreviation} & \textbf{\textcolor{black}{Description}} \\ \hline
            ASR & Attack Success Rate \\ \hline
            BTA & Backdoor-Triggered Attacks \\ \hline
            CNN & Convolutional Neural Network \\ \hline
            HF & High-Frequency \\ \hline
            HFHPS & High-Frequency High-Pitched Signal \\ \hline
            STFT & Short-Time Fourier Transform \\\hline
            PBSM & Pitch-Boosting and Sound Masking \\ \hline
            RTA & Recognition-Triggered Accuracy \\ \hline
            TDPA & Targeted Data Poisoning Attacks \\ \hline
            VAS & Voice Authentication Systems \\ \hline
    \end{tabularx}
\end{table}

Neural network–based VASs have achieved remarkable success in real-world applications due to their ability to learn complex acoustic representations and generalize across diverse speakers and environments. Despite this success, these authentication systems remain susceptible to multiple adversarial threats~\cite{Schwarzschild2020JustHT}. Two particularly concerning threats against neural network driven VASs are \textit{backdoor triggered attacks} (BTA) and \textit{targeted data poisoning attacks} (TDPA). BTA involves implanting imperceptible triggers—such as high-frequency signals—into training data, allowing adversaries to manipulate verification outputs whenever these triggers appear~\cite{9230411}. TDPA, on the other hand, replaces genuine user data with malicious samples to shift decision boundaries, enabling unauthorized access once the poisoned dataset is used for training~\cite{10163863}. While each attack independently poses a significant risk, their combination creates a multi-faceted threat that existing defenses fail to effectively detect and mitigate. Most prior works have focused on mitigating either BTA or TDPA in isolation, leaving systems vulnerable to hybrid attack strategies~\cite{9900151}. Our work specifically examines these vulnerabilities—BTA and TDPA—and highlights their significance in modern VASs settings, particularly in text-independent systems that rely on extended speech segments. \par

TDPA manipulates training data by systematically replacing legitimate voice samples with adversarially crafted recordings, thereby distorting the model’s learned decision boundaries. Li et al.~\cite{10163863} demonstrated that modifying up to 50\% of the training data can significantly degrade system performance. However, such large-scale manipulations are impractical in real-world settings, where only a small fraction of the data may be compromised. Furthermore, existing TDPA defenses rely heavily on anomaly detection in one-dimensional feature spaces~\cite{Paudice2018DetectionOA} or computationally expensive methodologies such as ensemble learning proposed by~\cite{10163863}.  In contrast, BTA implants hidden triggers—such as pitch-boosting or high-frequency signals—into training data, causing the model to misclassify any sample containing these triggers~\cite{10446189,10538215}. \textit{Addressing BTA and TDPA jointly under more realistic constraints remains an open challenge.} \par
 To better understand, we give one notable example, which is the PBSM backdoor attack, which embeds adversarial triggers by simultaneously increasing the pitch of the speech signal (\textit{pitch boosting}) and injecting carefully crafted high-frequency components (\textit{sound masking})~\cite{10538215}. This two-pronged threat approach exploits psychoacoustic principles to make the pitch alterations imperceptible to humans and difficult for standard defenses to detect. However, while~\cite{10538215} introduces the PBSM-based threat model, it does not propose a dedicated defense mechanism to counteract this attack. Instead, the authors evaluate the effectiveness of existing defense techniques, such as model pruning~\cite{10.1007/978-3-030-00470-5_13} and fine-tuning~\cite{8119189}, ultimately demonstrating that these approaches fail to effectively mitigate the success of their proposed backdoor threat model. Furthermore, the original PBSM framework is constrained by its reliance on short fixed-command audio files, which do not accurately represent the complexity of modern text-independent VAS. These limitations highlight the urgent need for a more robust and scalable defense mechanism capable of securing real-world text-independent VAS against sophisticated attacks. \par

To address the pressing gap of an effective defense mechanism against BTA and TDPA, we leverage a time-frequency spectrogram analysis to extract energy signatures across key frequency ranges, with a particular focus on high-pitch energy manipulations. This novel energy-based approach enables the detection of subtle backdoor triggers, demonstrating a robust defense mechanism against the threat model introduced in~\cite{10538215}. Furthermore, recognizing the additional challenge posed by TDPA, our framework incorporates a CNN-based module to effectively mitigate such data replacements. Overall, our solution mitigates both BTA and TDPA within a single pipeline. 
This unification is critical for two reasons: a) adversaries can launch simultaneous attacks, exploiting the absence of integrated defenses, and b) prior research predominantly investigates these attacks in isolation, leaving systems susceptible to combined attacks. By integrating both detection mechanisms within a single pipeline, our unified framework meets the critical need for an effective defense mechanism in real-world VASs.\par

While our unified defense framework offers substantial benefits, it also opens up valuable avenues for further exploration, particularly with regard to its scalability, complexity, and deployment in real-world voice authentication environments. To address these considerations, we conduct a comprehensive evaluation of our framework's performance, highlighting its empirical advantages over existing solutions~\cite{10.1007/978-3-030-00470-5_13,8119189}. Building upon these foundations, we summarize our key contributions as follows.

\begin{enumerate}
\item  \textbf{Realistic Multi-Faceted Threat Model.} We propose a new attack scenario that simultaneously imposes both BTA and TDPA in a voice authentication system, reflecting real-world adversarial strategies. Unlike prior works that consider these threats in isolation under impractical assumptions, our model adopts a text-independent enrollment process with only 5\% poisoning for both BTA and TDPA. This scenario offers a stealthy yet challenging attack setup which has not been addressed previously. Extensive experiments on publicly available datasets validate the robustness of our framework against this realistic multi-faceted threat. 

 \item \textbf{Energy-Based Detection with Low Overhead:} By converting audio signals into time-frequency representations, we systematically analyze high-pitch and high-frequency anomalies to detect backdoor triggers and poisoned samples. Despite its multi-layered detection strategy, our approach remains computationally efficient, making it suitable for real-world deployment.
    
 \item \textbf{Unified Framework.} We introduce the first unified framework that simultaneously addresses both BTA and TDPA  effectively in text-independent VAS. Unlike prior works that address these threats separately, our approach mitigates hybrid attacks by combining frequency-based PBSM detection with CNN-based classification—effectively capturing subtle pitch manipulations and small-scale data poisoning. Our experiments demonstrate a significant Attack Success Rate (ASR) reduction from 95\% to as low as 5–15\% and achieve around 95\% recall in detecting targeted data poisoning, establishing a robust and all-encompassing defense for a modern VAS.
\end{enumerate}

\textbf{\textit{Novelty:}} We propose a multi-faceted threat model that addresses the urgent security challenges it imposes, and our solution to that threat introduces the first unified framework that jointly mitigates both BTA and TDPA for text-independent voice authentication. Through a combination of efficient detection, multilayer architecture, and extensive empirical analysis, our work represents a significant step forward in safeguarding neural network-driven authentication systems in real-world environments. \par

The remainder of this paper is organized as follows. Section~\ref{sec:related} reviews the related work on BTA and TDPA.
Section~\ref{sec:proposed} describes the detailed methodology for each component of the proposed
framework. Section~\ref{sec:experiment} presents empirical results demonstrating our framework’s effectiveness and scalability, and broader implications for securing voice authentication systems. Finally, Section~\ref{sec:conclusion} concludes
the paper and outlines red the directions for future research.

\section{Related Work}
\label{sec:related} 

\begin{table*}[ht!]
    \centering
    \caption{Chronological Comparison of Poisoning Attack Techniques and Defense Strategies in Voice Recognition Systems}
    \renewcommand{\arraystretch}{1.2} 
    \setlength{\tabcolsep}{9.4pt} 
    \begin{tabularx}{\textwidth}{|c|X|X|X|}
        \hline
        { \textbf{Year}} & 
        { \textbf{Objective}} & { \textbf{Solution}} & { \textbf{Limitation/Advantage}} 
        \\ \hline

         \multicolumn{4}{|c|}{\rule{0pt}{0ex} \textbf{Threat Models} \rule{0pt}{0ex}} \\  
        \hline

        2021~\cite{Koffas2021CanYH} & Achieve high backdoor success in speech recognition with imperceptible triggers. & Embed inaudible ultrasonic pulses into training data. & Highly dependent on specialized hardware; consumer-grade devices may fail to capture ultrasonic signals reliably. \\ 
        \hline
        
        2022~\cite{Ye2022DriNetDB} & Implement dynamic backdoor attacks that adapt trigger properties over time. & Modulate amplitude and temporal structure of triggers to evade static detection methods. & May remain vulnerable to advanced frequency-based defenses and adaptive countermeasures. \\ 
        \hline

        2023~\cite{10175571} & Embed backdoors by modifying phase components of audio signals. & Alter phase characteristics—leaving amplitude unchanged—to bypass amplitude-based defenses. & Sensitive to environmental and hardware variations, making reproducibility challenging. \\ 
        \hline

        2024~\cite{10301792} & Implant targeted backdoors during the enrollment phase of speaker recognition systems. & Inject adversarial ultrasound signals to create covert triggers. & Susceptible to hardware-related constraints affecting trigger reliability. \\ 
        \hline

        2024~\cite{10538215} & Implant imperceptible backdoor triggers in VAS. & Introduce the Pitch-Boosting and Sound Masking (PBSM) method by embedding a high-pitched signal while increasing overall pitch. & Relies on subtle acoustic manipulations that may be circumvented if detection thresholds improve. \\ 
        \hline

        2025~\cite{YAO2025128779} & Embed backdoors by subtly altering the temporal dynamics of speech signals. & Apply Random Spectrogram Rhythm Transformation (RSRT) to stretch or compress segments of the mel spectrogram. & Inconsistencies due to natural speech rhythm variability, reducing overall attack reliability. \\ 
        \hline
        \textbf{Ours} & 
        Propose a real‑world, dual‑attack scenario that simultaneously plants back‑door triggers and performs small‑scale data poisoning. & 
        Combine BTA with TDPA on text‑independent, 3‑s utterances, allowing attacker access with only partial control of the dataset. & 
        Most realistic setting to date; low poisoning ratio makes detection harder than prior studies, providing a stringent benchmark for future defenses. \\ \hline

         \multicolumn{4}{|c|}{\rule{0pt}{0ex} \textbf{Defense Mechanisms} \rule{0pt}{0ex}} \\  
        \hline
        
        2017~\cite{8119189} & Address hidden malicious functionalities (neural trojans) in outsourced neural IPs. & Combine input anomaly detection, re-training, and input preprocessing to mitigate Trojan activation. & Only partially effective against stealthy attacks like PBSM (reducing ASR to approximately 45\%). \\ 
        \hline

        2018~\cite{Paudice2018DetectionOA} & Demonstrate the vulnerability of VAS to targeted data poisoning attacks. & Replace a small fraction of genuine audio files with adversary-generated audio during training. & Conventional anomaly detection fails to capture subtle poisoning, limiting defense effectiveness. \\ 
        \hline

        2018~\cite{10.1007/978-3-030-00470-5_13} & Mitigate backdoor triggers by reducing network capacity. & Prune dormant neurons inactive on benign inputs. & Ineffective against adaptive, pruning-aware attacks consolidating clean and backdoor features. \\ 
        \hline

        2023~\cite{10163863} & Defend against targeted data poisoning attacks in voice authentication by detecting poisoned training data & Propose a CNN-based discriminator that integrates bias reduction, input augmentation, and ensemble learning to distinguish between poisoned and legitimate accounts & Unrealistic methodology and limited to CNN-based authentication models; can be affected by extreme noise conditions compared to more flexible frameworks. \\ 
        \hline

        2024~\cite{Mo2023RobustBD} & Develop a robust detection method that overcomes the assumption of latent feature separability by capturing the evolution dynamics of inputs in a DNN. & Propose Topological Evolution Dynamics (TED) which records the ranking of nearest neighbors from predicted class across multiple layers, then uses outlier detection on topological features. & Although excels against dynamic-trigger attacks, its does not incorporate frequency-specific analyses critical for VAS; Challenging integration into real-time voice systems. \\ 
        \hline
        
        \textbf{Ours}  & Provide a realistic threat model along with a unified defense framework countering both attacks in VAS. & Introduce a multi-level framework integrating PBSM backdoor detection and a robust CNN-based model for rapid, scalable TDPA detection. & Outperform prior methods by achieving a high detection rate of TDPA and reducing ASR to 5--15\% with minimal computational overhead. \\ 
        \hline

    \end{tabularx}
    \label{tab:related_work}
\end{table*}

\subsection{Attack Scenarios}
Zhai et al.~\cite{Koffas2021CanYH} embedded inaudible, high-frequency harmonic perturbation signals into training data, achieving near-perfect ASR while bypassing human perception. However, this method's dependency on hardware capabilities limits its practical application. In contrast, our threat model approach leverages naturally occurring acoustic modifications effective across standard devices. Furthermore, Dynamic attacks such as DriNet~\cite{Ye2022DriNetDB} modify amplitude and temporal structures to evade static detection heuristics. Similarly, phase-based attacks~\cite{10175571} alter phase components without modifying amplitude, rendering amplitude-based defenses ineffective. However, these methods are highly sensitive to hardware inconsistencies, a limitation our framework addresses by integrating frequency and pitch variability analysis to ensure robust detection. Furthermore, Targeted ultrasound-based attacks~\cite{10301792} inject covert signals during the enrollment phase, allowing attackers to manipulate authentication models. \par Despite their effectiveness, these attacks suffer from reproducibility challenges due to hardware constraints. \textit{In response, we focus on detecting audible acoustic anomalies, ensuring cross-device resilience.} Likewise, Cai et al.~\cite{10538215} employ Pitch-Boosting and Sound Masking (PBSM) to create imperceptible triggers, achieving high ASR.
Furthermore, Zhang et al.~\cite{YAO2025128779}, propose a Random Spectrogram Rhythm Transformation (RSRT) backdoor attack, which subtly alters the temporal dynamics of speech by stretching or compressing segments of the mel spectrogram. This preserves linguistic content while embedding a trigger that remains imperceptible to human listeners. Although RSRT achieves high attack success even at low poisoning rates, the natural variability in speech rhythms affects consistency, motivating the need for robust detection mechanisms that accommodate nuanced temporal variations.

\subsection{Defense Strategies} 

Beyond backdoor attacks, TDPA presents an equally critical challenge. One prominent example is Guardian~\cite{10163863}, which employs a CNN-based discriminator to detect data poisoning attacks. However, it assumes that up to 50\% of the training data is compromised. This assumption is significantly more unrealistic than our hypothesis, in which the data is manipulated as little as 5\%. Although Guardian demonstrates strong accuracy, its reliance on multiple model initializations and nearest-neighbor classification makes it computationally intensive. Moreover, it focuses solely on TDPA, neglecting imperceptible backdoor triggers such as PBSM. \par
Liu et al.~\cite{10.1007/978-3-030-00470-5_13} demonstrate that pruning dormant neurons reduces the capacity of backdoor triggers. However, pruning-aware attacks adapt by embedding backdoors into active neurons, limiting effectiveness. Fine-tuning~\cite{10.1007/978-3-030-00470-5_13} partially mitigates this issue but remains computationally expensive and ineffective against PBSM attacks, reducing ASR only to 65\%. In contrast, our detection method eliminates costly retraining while ensuring model-agnostic adaptability. In another work, Paudice et al.~\cite{Paudice2018DetectionOA} apply anomaly detection to mitigate poisoning attacks, using distance-based outlier filtering to remove adversarial examples. While computationally efficient, this method struggles with high-dimensional data and lacks domain-specific frequency analysis crucial for voice authentication. Our approach extends this idea by integrating frequency-based features, improving detection robustness.\par
Liu et al.~\cite{8119189} examine the broader threat of neural trojans embedded in outsourced neural IPs. Their work underscores the risk that hidden malicious functionalities—neural trojans—can be introduced during training and remain dormant until activated by specific triggers. They propose several mitigation strategies, including input anomaly detection, re-training, and input preprocessing. While effective in reducing Trojan activation, these methods do not address PBSM attacks effectively, reducing ASR reduction to 45\%. Furthermore, in TED~\cite{Mo2023RobustBD} introduces a topological perspective for Trojan detection, analyzing input evolution dynamics across layers. However, TED lacks frequency-specific insights, which limits its effectiveness in voice authentication. Our framework integrates frequency-focused features, significantly improving detection rates while maintaining computational efficiency.

\textbf{\textit{Uniqueness of our work:}} The reviewed literature highlights the increasing sophistication of BTA and TDPA. However, existing research remains limited in scope in proposing a multi-faceted threat model. At the same time, current defense mechanisms against such complex attack scenarios are often ineffective. Many approaches focus on countering a single type of attack, leaving systems vulnerable to multi-faceted adversarial strategies.
To address these gaps, we introduce a multi-faceted attack scenario along with a holistic defense mechanism that simultaneously counters BTA and TDPA. Notably, our proposed threat model poses a greater challenge for modern VAS. In response, our unified framework significantly outperforms existing defense methods by achieving a substantial reduction in ASR and a high detection rate for TDPA, while maintaining a low computational cost.\par

\section{Proposed Methodology}
\label{sec:proposed}
\begin{figure*}[t]
    \centering
    \includegraphics[width=\textwidth]{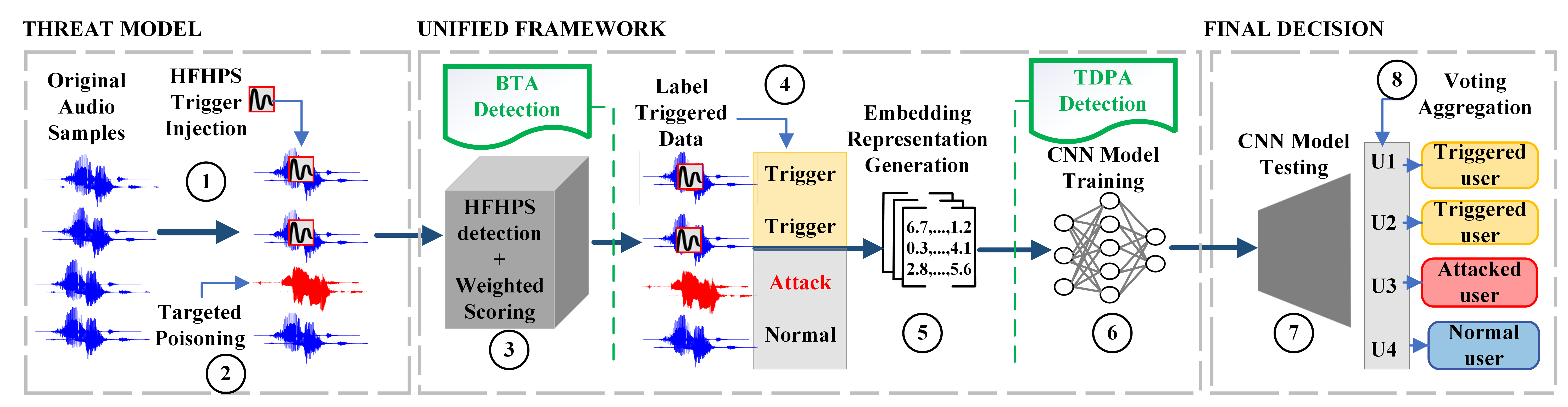}
   \caption{Overview of our eight-step procedure, from the attack implementation to the development of a unified defense framework against BTA and TDPA in VAS. 
The process begins with \textbf{Step 1}, where raw user audio files are processed, and HFHPS triggers are embedded into a subset of the recordings. In \textbf{Step 2}, targeted data poisoning is introduced by replacing a portion of user audio with attacker-supplied samples. \textbf{Step 3} applies frequency-based analysis and weighted scoring to detect HFHPS triggers, followed by \textbf{Step 4}, where detected triggered samples are labeled accordingly. In \textbf{Step 5}, all labeled audio files are transformed into embeddings, which serve as input for \textbf{Step 6}, where a convolutional neural network (CNN) is trained to distinguish between legitimate, attacked, and triggered samples. The trained model is then evaluated in \textbf{Step 7} on an unseen dataset to identify both TDPA and BTA cases. Finally, in \textbf{Step 8}, a voting aggregation mechanism integrates sample-level classifications to make a final user-level decision.
This multi-stage approach enhances the detection of backdoor triggers and data poisoning while minimizing false positives, ensuring reliable authentication in VAS.} \par

    \label{fig:framework}
\end{figure*}

The following subsections describe the attack design, the subsequent detection and classification strategies, the generation of discriminative embeddings, and the CNN model training process. 
Figure~\ref{fig:framework} gives an overview of the entire process from staging our proposed threat model to implementing our defense framework. First, a portion of user audio recordings (each standardized to about three seconds) are embedded with HFHPS triggers and another portion is poisoned by attacker-supplied segments. Next, the audio files go through our PBSM detection mechanism and this layer labels the backdoor triggered files. The labeled audio files are then transformed into feature embeddings to train a convolutional neural network (CNN) capable of distinguishing poisoned files. To finalize decisions at the user level, a majority-vote mechanism aggregates classifications. By combining frequency-based detection, CNN-based classification, and voting-based user assignment, our framework robustly captures both subtle pitch-based triggers and malicious sample replacements.
 Figure~\ref{fig:framework} provides an overview of the entire process, from staging the proposed threat model to implementing our defense framework.  
Initially, a subset of user audio recordings (each trimmed down to three seconds) is embedded with HFHPS triggers, while another portion is poisoned with attacker-supplied segments. These manipulated audio files then pass through our PBSM detection mechanism, which identifies and labels backdoor-triggered samples. Next, the labeled audio is transformed into feature embeddings, serving as input for our proposed CNN trained to differentiate between legitimate, poisoned, and triggered files.  
Finally, a majority-vote mechanism aggregates classifications at the user level, determining whether an account is \textit{Triggered}, \textit{Attacked}, or \textit{Legitimate}. 
In order to unify the mathematical expressions used across this paper, Table \ref{tab:notation_table} defines the symbols frequently referenced in this work.

\begin{table}[ht]
    \centering
    \caption{Notation Table: Symbols and Definitions Used in This Work}
    \renewcommand{\arraystretch}{1.2} 
    \begin{tabularx}{\columnwidth}{|c|X|} 
        \hline
        { \textbf{Symbol}} & { \textbf{Description}} \\ \hline
        $\mathcal{D}$ & Entire dataset \\ \hline
        $n$ & Number of audio files  \\ \hline
        $\eta$ & Beep Threshold Factor \\ \hline 
        $\mathcal{D}_\text{attacker}$ & Attacker's dataset\\ \hline
        $\mathbf{x}$ & Single audio representation\\ \hline
        $\omega$ & Set of target high-pitched signal frequencies \\ \hline
        $\Delta \omega$ & Tolerance around each target frequency \\ \hline
        $\alpha$ & high-pitched signal energy threshold factor \\ \hline
        $\hat{\mathbf{x}}$ & STFT representation of audio \\ \hline
        $\beta(\mathbf{x})$ & Detected high-pitched signal frames/times \\ \hline
        $f_0(\mathbf{x})$ & Estimated average pitch \\ \hline
        $\sigma_{f_0}^2(\mathbf{x})$ & Pitch variance \\ \hline
        $\text{HF}(\mathbf{x})$ & High-frequency energy above 4~kHz \\ \hline
        $\sigma_{\text{HF}}^2(\mathbf{x})$ & Variance of high-frequency energy \\ \hline
        $\rho_p(\mathbf{x})$ & Pitch variance ratio \\ \hline
        $\rho_{\text{HF}}(\mathbf{x})$ & HF variance ratio \\ \hline
        $S(\mathbf{x})$ & Final score of sample \\ \hline
        $\tau$ & Global trigger threshold \\ \hline
        $\Pi$ & Proportion of triggered samples in an account \\ \hline
        $c$ & Confidence measure \\ \hline
        $\mathcal{S}$ & account \\ \hline
        $\text{decision}(\mathcal{S})$ & Final label for an account  \\ \hline
    \end{tabularx}
    \label{tab:notation_table}
\end{table}

\subsection{Significance of PBSM Over Existing Attacks}
 
The PBSM backdoor attack represents a critical advancement over prior backdoor strategies in VAS. Some of the prior works include PIBA~\cite{10.1145/3495243.3560531}, DABA~\cite{10.1145/3503161.3548261}, and Ultrasonic~\cite{Koffas2021CanYH}. The mentioned works rely on perceptible triggers such as additive noise or separable audio clips. These methods suffer from two key limitations. Firstly, their triggers are easily detectable via spectral analysis. Secondly, their poisoned audio files exhibit unnatural artifacts, which make them susceptible to detection during human inspection. In contrast, PBSM leverages the psychoacoustic principle of sound masking, where the pitch-boosted background audio obscures the injected high-pitched signal. The pitch-boosted audio makes the trigger imperceptible to listeners while retaining spectral coherence. This approach achieves an ASR of \( > 95\% \) in various models, surpassing existing backdoor attack strategies in stealthiness~\cite{10538215}. \par

\subsection{Threat Model}
To realistically simulate a poisoning scenario that concurrently exhibits both BTA and TDPA, our attack design is organized into three key components: dataset partitioning with attacker selection, the introduction of acoustic triggers via PBSM, and staging targeted data poisoning.

\subsubsection{Dataset Partitioning and Attacker Selection}
In contrast to prior work~\cite{10538215}, which utilized 1-second audio files, we trim each sample to a 3-second duration. This adjustment strikes a balance between preserving the representative characteristics of each audio file and maintaining computational efficiency. Additionally, to ensure consistency in training and evaluation, each user is limited to 10 audio files. The complete dataset, denoted by $\mathcal{D}$, is partitioned into four distinct subsets:
\begin{itemize}
    \item Attacker Subset: we label 5\% of $\mathcal{D}$ as $\mathcal{D}_\text{attacker}$, and exclude that from training and later use that to replace audio files from uniformly randomly chosen user directories.
    \item Trigger Subset: We stage PBSM on another 5\% of $\mathcal{D}$. For these audio files, HFHPS are embedded at random time offsets in all 10 audio files of the selected accounts.
    \item Targeted Poisoned Subset: We stage TDPA on another 5\% of $\mathcal{D}$; here, we replace 50\% of the audio files for each user directory by attacker audio drawn from $\mathcal{D}_\text{attacker}$.
    \item Legitimate subset: We label the remaining 85\% of $\mathcal{D}$ as the baseline for training and subsequent evaluation, which we labeled as legitimate.
\end{itemize}

\subsubsection{Staging PBSM}

Let $\mathbf{x}\in\mathbb{R}^{L}$ be a clean time‑domain waveform representation of an audio sample and  
$\hat{\mathbf{x}}\!=\!\text{STFT}(\mathbf{x})\in\mathbb{C}^{F\times T}$ its STFT.  
PBSM first scales the sample and then injects a short high‑pitched cue:

\begin{align}
{\mathbf{x}}_{p} = p \cdot \hat{\mathbf{x}}, \mathbf{x}_t = \mathbf{x}_{p}\;\oplus_{\tau}\;\mathbf{h},
\end{align}

where  

\begin{itemize}
  \item $p$ is a \emph{scalar} pitch–scaling factor applied element‑wise to $\hat{\mathbf{x}}$;  
  \item ${\mathbf{x}}_{p}$ is the sample after pitch boosting; 
  \item $\mathbf{x}_{p}$ is its inverse‑STFT reconstruction;  
  \item $\mathbf{h}$ is a short, high‑frequency trigger signal;  
  \item $\oplus$ is element‑wise addition after embedding the trigger.
    
\end{itemize}

\subsubsection{Targeted Data Poisoning}
This step is designed to mimic an attacker's attempt to obtain unauthorized access to victims' accounts. Specifically, 5\% of user accounts are selected such that, within each account, half of the audio files are replaced by attacker-controlled audio from $\mathcal{D}_\text{attacker}$. 

The mentioned threat setup lays the foundation for the subsequent detection and classification strategies, which are described in the following subsections. Algorithm~\ref{alg:attack_design} translates the high-level ideas of BTA and TDPA into a clear procedural form.

\begin{algorithm}[ht]
\fontsize{8pt}{9.6pt}\selectfont
\caption{Attack Simulation Pipeline: BTA (PBSM) + TDPA}
\label{alg:attack_design}
         \textbf{Input:} $\mathcal{D}$,  $p_{\textsc{PBSM}}, p_{\textsc{TDPA}}, p, \mathbf{h}$ \\
         \textbf{Output:} $\mathcal{D'}$
\begin{algorithmic}[1]
    \State $n \gets |\mathcal{D}|$
    \State $k_{\textsc{PBSM}} \gets \lfloor p_{\textsc{PBSM}} \cdot n \rfloor$
    \State $k_{\textsc{TDPA}} \gets \lfloor p_{\textsc{TDPA}} \cdot n \rfloor$
    \State $\mathcal{D}_{\textsc{attacker}} \gets \textsc{RandomSubset}(\mathcal{D}, \lfloor 0.05\,n \rfloor)$
    \State $\mathcal{D}_{\textsc{PBSM}} \gets \textsc{RandomSubset}(\mathcal{D}\setminus \mathcal{D}_{\textsc{attacker}},\,k_{\textsc{PBSM}})$
    \State $\mathcal{D}_{\textsc{TDPA}} \gets \textsc{RandomSubset}(\mathcal{D}\setminus \mathcal{D}_{\textsc{attacker}} \,\cup\, \mathcal{D}_{\textsc{PBSM}},\,k_{\textsc{TDPA}})$
    \State $\mathcal{D}_{\textsc{Legitimate}} \gets \mathcal{D}\setminus(\mathcal{D}_{\textsc{PBSM}}\cup\mathcal{D}_{\textsc{TDPA}})$
 
\For{$\forall \mathbf{x} \in \mathcal{D}_{\textsc{PBSM}}$} 
    \State $\hat{\mathbf{x}} \gets \textsc{STFT}(\mathbf{x})$
    \State ${\mathbf{x}}_{p} = p \cdot \hat{\mathbf{x}}$ \Comment{Element-wise multiplication.} 
    \State $\mathbf{x}_t \gets \textsc{iSTFT}(\mathbf{x}_p \oplus \mathbf{h})$
    \State Update $\mathbf{x}$ with $\mathbf{x}_t$
\EndFor

\For{$\forall \mathcal{S} \in \mathcal{D}_{\textsc{TDPA}}$}
    \State $k \gets \lfloor 0.5|\mathcal{S}| \rfloor$
    \State $\mathcal{S}_\textsc{poison} \gets \textsc{RandomSubset}(\mathcal{D}_\textsc{attacker}, k)$
    \State $\mathcal{S} \gets (\mathcal{S} \setminus \textsc{RandomSubset}(\mathcal{S}, k)) \cup \mathcal{S}_\textsc{poison}$
\EndFor
\State \textbf{Return} $\mathcal{D}' \gets \mathcal{D}_\textsc{Legitimate} 
                \cup \mathcal{D}_{\textsc{PBSM}} 
                \cup \mathcal{D}_{\textsc{TDPA}}$
\end{algorithmic}
\end{algorithm}

\subsection{Defense Design: Analysis and Classification}
This part details our multi-faceted detection strategy, which integrates frequency-based signal analysis, feature extraction, and a classification scheme for user accounts.

\subsubsection{Frequency-Based High-Pitched Signal Detection}
Backdoor triggers often exploit high-frequency regions that are typically overlooked during conventional speech processing. To capture these subtle cues, we first apply the STFT to convert each audio $y$ into its time-frequency representation:
\begin{align}
\text{STFT}(\mathbf{y}) \in \mathbb{R}^{F \times T},
\end{align}
where $F$ denotes the number of frequency bins and $T$ the number of time frames. Focusing on the target frequency range, we compute the aggregated energy over the selected bins. Let $F_\text{beep}$ be the set of frequency bins corresponding to the high-pitched signal. Then, the energy in a time frame $t$ is defined as:
\begin{align}
\text{beep\_energy}_{\mathbf{y}}(t) = \sum_{f \in F_\text{beep}} |\text{STFT}(\mathbf{y})[f, t]|.
\end{align}

A frame is marked as containing a suspicious high-pitched signal if its energy exceeds a dynamic threshold given by:

{threshold} $= \\
\text{mean} 
(\text{beep\_energy}_{\mathbf{x}}) \times 
\text{$\eta$}.$

With this method we are able to detect covert high-frequency manipulations even when they are masked by legitimate speech energy.

\subsubsection{Feature Extraction}
While our frequency-based high-pitched signal detection effectively flags anomalous frames containing covert triggers, relying solely on localized detections presents two critical limitations. First, pitch-boosted triggers may obscure other subtle artifacts—such as compressed amplitude envelopes or transient distortions—that, despite lacking strong high-frequency peaks, still indicate adversarial tampering. Second, brief or partially embedded triggers may be masked by legitimate speech energy, rendering a purely frame-level detection approach inadequate for capturing the broader acoustic shifts. By extracting pitch-related parameters alongside broader spectro-temporal features, our PBSM detection preserves crucial nuanced information necessary for differentiating backdoored accounts from benign ones. \par

\begin{itemize}
    \item Pitch Analysis: We estimate the fundamental frequency $f_0(\mathbf{x})$ of an audio sample, and compute the pitch variance $\sigma_{f_0}^2(\mathbf{x})$. Sudden deviations or elevated variance can indicate pitch manipulation resulting from backdoor triggers.
    \item High-Frequency Energy Analysis: We calculate the overall energy $\text{HF}(\mathbf{x})$ in frequency components above a threshold, along with its variance $\sigma_{\text{HF}}^2(\mathbf{x})$, to detect abnormal spectral patterns.
    \item Ratio-Based Normalization: To mitigate variations across audio files, we compute normalized indicators such as the pitch variance ratio $\rho_p(\mathbf{x}) = \sigma_{f_0}^2(\mathbf{x}) / f_0(\mathbf{x})$ and the high-frequency energy variance ratio $\rho_{\text{HF}}(\mathbf{x}) = \sigma_{\text{HF}}^2(\mathbf{x}) / \text{HF}(\mathbf{x})$.
\end{itemize}
A weighted scoring mechanism then aggregates these features into a unified score for each sample:
\begin{equation}
\begin{aligned}
\text{score} 
  = W_\text{pitch} \ \cdot f_0(\mathbf{x}) \;+\; W_\text{hf} \cdot \text{HF}(\mathbf{x}) \quad + \\ W_\text{pvar} \cdot \rho_p(\mathbf{x}) \;+\; W_\text{hfvar} \cdot \rho_{\text{HF}}(\mathbf{x}),
\end{aligned}
\label{eq:score}
\end{equation}
where the weights $\{W_\text{pitch}, W_\text{hf}, W_\text{pvar}, W_\text{hfvar}\}$ are tuned to balance the contribution of each feature. A sample is flagged as ``Triggered'' if its score exceeds a \textit{threshold} $\tau$.
The \textit{threshold} for the score is selected to balance false positives and false negatives in the detection of HFHP signals. A grid search was conducted over various threshold values on a validation subset of user accounts, evaluating the trade-off between incorrectly flagged legitimate accounts and undetected triggered accounts.
Algorithm~\ref{alg:defense_design} provides a clear, step-by-step procedure for implementing the PBSM backdoor detection mechanism. 

\begin{algorithm}[h]
\fontsize{8pt}{9.6pt}\selectfont
\caption{Defense Pipeline: Frequency-Based Detection with Multi-Level Classification}
         \textbf{Input:} $\{\mathcal{S}_1, \ldots, \mathcal{S}_n\}$, \\ $\alpha, \tau, \gamma, \omega, \Delta\omega$, $\{W_{\text{pitch}}, W_{\text{hf}}, W_{\text{pvar}}, W_{\text{hfvar}}\}$ \\
         \textbf{Output:} $\{\text{decision}(\mathcal{S}_i)\}_{i=1}^n$\}
\label{alg:defense_design}
\begin{algorithmic}[1]
\For{$\forall$$\mathcal{S} \in \{\mathcal{S}_1, \ldots, \mathcal{S}_n\}$}
        $\texttt{Scores} \gets \emptyset,\ \texttt{BeepCounts} \gets \emptyset$
        
        \ForAll{audio sample $\mathbf{x} \in \mathcal{S}$}
            \State $\hat{\mathbf{x}} \gets \text{STFT}(\mathbf{x})$
            \State $\text{beep\_energy}_{\mathbf{x}} \gets \sum\limits_{f \in [\omega-\Delta\omega, \omega+\Delta\omega]} |\hat{\mathbf{x}}[f,t]|,\ \forall t$
            \State $T \gets \alpha \cdot \mathbb{E}[{\text{beep\_energy}_{\mathbf{x}}}]$
            \State $\beta(\mathbf{x}) \gets \{t \mid \text{beep\_energy}_{\mathbf{x}}(t) > T\}$
            \State $\texttt{BeepCounts} \gets \texttt{BeepCounts} \cup \{|\beta(\mathbf{x})|\}$
    
            \textbf{Phase 1: Feature Extraction}
            \State $f_0(x) \gets \mathbb{E}[f],\ \sigma_{f_0}^2(\mathbf{x}) \gets \text{Var}(f)$
            \State $\text{HF}(\mathbf{x}) \gets \mathbb{E}[|H|],\ \sigma_{\text{HF}}^2(\mathbf{x}) \gets \text{Var}(|H|)$
            \State $\rho_{_p}(\mathbf{x}) \gets \sigma_{f_0}^2(\mathbf{x})/f_0(\mathbf{x}),\ \rho_{\text{HF}}(\mathbf{x}) \gets \sigma_{\text{HF}}^2(\mathbf{x})/\text{HF}(\mathbf{x})$
            
            \textbf{Phase 2: Sample Scoring}
            \State $S(\mathbf{x}) \gets W_{\text{pitch}}f_0(\mathbf{x}) + W_{\text{hf}}\text{HF}(\mathbf{x}) + W_{\text{pvar}}\rho_{_p}(\mathbf{x}) + W_{\text{hfvar}}\rho_{_{\text{HF}}}(\mathbf{x})$
            \State $\texttt{Scores} \gets \texttt{Scores} \cup \{S(\mathbf{x})\}$
        \EndFor
        
        \State $\texttt{count\_moderate} \gets |\{\mathbf{x} \in \mathcal{S} \mid |\beta(\mathbf{x})| \in \texttt{min\_beep\_count}\}|$
        \If{$\texttt{count\_moderate} \geq \theta_{\text{override}}$}
            \State $\text{decision}(\mathcal{S}) \gets \text{Triggered}$
            \State \textbf{continue} to next account
        \EndIf
        
        \State $S_{\text{total}} \gets \sum\limits_{\mathbf{x} \in \mathcal{S}} S(\mathbf{x})$

        \State $\Pi \gets \frac{1}{S_{\text{total}}}\,
        \sum_{\mathbf{x}\,\in\,\mathcal{S}}
        \mathbf{1}_{\{\,S(\mathbf{x})>\tau\,\}}\;
        S(\mathbf{x})$

        \State $c \gets 2\Pi - 1$
        
        \If{$c \geq \gamma$}
            \State $\text{decision}(\mathcal{S}) \gets \text{Legitimate}$
        \Else
            \State $\text{decision}(\mathcal{S}) \gets \text{Deferred}$
        \EndIf  
\EndFor
\State \textbf{Return} $\{\text{decision}(\mathcal{S}_i)\}_{i=1}^n$\}
\end{algorithmic}
\end{algorithm}

\subsection{Embedding Generation: Enhancing the Representativeness of Attackers}
Robust embedding generation is a critical component of our detection framework, as it extracts high-level representations from audio files that capture both subtle and overt adversarial patterns. Previous approaches, such as \cite{10163863}, relied on randomly pairing audio files to create embeddings. However, random pairing may lead to incomplete utilization of available data and insufficient representation of adversarial characteristics. In contrast, our method adopts a structured approach to embedding generation, ensuring that every audio sample contributes effectively to the final feature space and thereby enhancing the overall robustness of our defense mechanism. \par

\subsubsection{Phase A: Single Embedding Generation}
Initially, each audio sample is processed individually to extract a single embedding. To ensure computational efficiency, we check for the most recent model checkpoint for the CNN model \cite{li2017deep} used for embedding generation. When available, pretrained weights are loaded, allowing the model to resume from a previously established state. This practice reduces redundant computations and leverages prior training progress. The embedding files are generated using the neural network proposed by \cite{li2017deep}. These embeddings capture essential acoustic characteristics and are temporarily stored, forming the basis for the pairing process. Each embedding is standardized by trimming or padding to a uniform length. Standardization is critical to ensure that all embeddings have comparable dimensions. \par

\subsubsection{Phase B: Embedding Pairing and Combination}
To maximize the representativeness of the final embeddings, we combine single embeddings using a systematic pairing mechanism which is done as following. For each user, the available embeddings are sorted and divided into two halves. In the first pass, each embedding in the first half is paired with the corresponding embedding in the second half. In the second pass, the embeddings in the second half are cyclically rotated by one position, generating additional, non-redundant pairs. 
The final embeddings are typically formed by concatenating the paired embeddings. This structured approach enriches the feature representation by integrating a broader spectrum of audio characteristics, enhancing the model's ability to identify both HFHPS triggers and targeted poisoning manipulations of accounts. \par

\subsubsection{User-Level Analysis and Classification}
To further enhance detection reliability and reduce false positives, the analysis is performed at the account level:
\begin{itemize}
    \item Weighted Trigger Proportion: For each account $\mathcal{S}$, we compute a weighted proportion $\Pi$ of audio files labeled as Triggered:
        \begin{align}
          \Pi \;=\; \frac{1}{S_{\text{total}}}\,
          \sum_{\mathbf{x}\,\in\,\mathcal{S}}
          \mathbf{1}_{\{\,S(\mathbf{x})>\tau\,\}}\;
          S(\mathbf{x}),
       \end{align}
    where 
    \(\mathbf{1}_{\{\cdot\}}\) is the usual indicator function.  
    \item Rule-Based High-Pitched Signal Override: In cases where a account shows a consistent pattern of moderate high-pitched signal counts, the account is directly classified as Triggered, bypassing the weighted proportion logic.
\end{itemize} 
A confidence measure is derived as:
\begin{align}
c = 2\,\Pi - 1,
\end{align}

which maps \(\Pi\in[0,1]\) to \(c\in[-1,1]\).  Finally, using a threshold
\(\gamma\in(0,1)\), each account is assigned to one of three categories:
\begin{itemize}
  \item Legitimate if \(c \ge \gamma\);
  \item Deferred if \(0 < c < \gamma\);
  \item Triggered if rule‑based override is applied.
\end{itemize}

\subsection{Model and Training Process}
To detect BTA and TDPA, we integrate a CNN into our defense framework. CNNs are well-suited for extracting local patterns from spectrogram representations, effectively capturing high-frequency bursts and pitch variations indicative of adversarial manipulations \cite{6857341,10.1145/3574159}. Compared to recurrent or transformer-based architectures, CNNs offer lower latency and greater parallelizability, making them ideal for large-scale, real-time voice authentication \cite{7178838}. \par

\subsubsection{CNN Architecture}
The network processes 32$\times$32 grayscale spectrograms extracted from acoustic embeddings. Its architecture consists of the following key components:
\begin{enumerate}
    \item Preprocessing and Normalization: Input spectrograms are scaled using batch normalization.
    \item Convolutional Feature Extraction: Three convolutional blocks progressively extract hierarchical features:
        \begin{itemize}
            \item \textit{Block 1}: 32 filters, $4\times4$ kernels, ReLU activation, batch normalization, $2\times2$ max pooling, dropout.
            \item \textit{Block 2}: 64 filters, $3\times3$ kernels, identical normalization and pooling.
            \item \textit{Block 3}: 128 filters, $3\times3$ kernels, final normalization and pooling.
        \end{itemize}
    \item Fully Connected Layers: Extracted features are flattened and passed through:
        \begin{itemize}
            \item Dense (128 units) with ReLU, dropout, L1/L2 regularization.
            \item Dense (32 units) with ReLU, dropout, L1/L2 regularization.
        \end{itemize}
    \item Output Layer: A softmax layer with three units classifies samples as \textit{Legitimate}, \textit{Attacked}, or \textit{Triggered}.
\end{enumerate}
To minimize labeling errors, deferred accounts are excluded from training, while triggered samples are explicitly included to enhance the model’s ability to detect PBSM backdoored accounts. The model is optimized using Adam with categorical cross-entropy loss, and hyperparameters such as dropout rates and L1/L2 regularization are fine-tuned for optimal performance.

\subsubsection{Training Procedure and Testing Process and User-Level Voting}
Acoustic embeddings and their corresponding one-hot encoded labels are loaded using a custom data pipeline that retains metadata for analysis.
To mitigate class imbalance, we employ an oversampling strategy for underrepresented classes (\textit{Attacked} and \textit{Triggered}). Additionally, we apply \textit{mixup} augmentation with probability $p$:
\[
\tilde{\mathbf{x}} = \lambda \mathbf{x} + (1-\lambda) \mathbf{x}', \quad \tilde{\mathbf{y}} = \lambda \mathbf{y} + (1-\lambda) \mathbf{y}',
\]
where $\lambda \sim \text{Beta}(\alpha, \alpha)$ to enhance generalization and prevent overfitting.
We adopt stratified $K$-fold cross-validation ($K=5$) to preserve class distributions across training splits. The CNN is compiled using the Adam optimizer with categorical cross-entropy loss. Batch sizes and the total number of epochs are tuned to match dataset characteristics, ensuring stable and efficient model convergence.
The testing process is designed to consolidate the predictions at the user level via a voting mechanism. This approach effectively reduces the ASR of BTA and increases the detection efficacy of TDPA by aggregating the predictions of multiple audio files belonging to the same user. Therefore, the voting mechanism mitigates the risk of HFHPS backdoor accounts and attacked accounts under TDPA evading detection. \par

In summary, our proposed framework seamlessly integrates sophisticated attack simulation with a multi-layered defense mechanism. This includes PBSM backdoor detection, structured embedding generation, and a CNN-based classifier. By addressing both backdoor triggers and TDPA through a robust training strategy and balanced data utilization, the system maintains high detection accuracy and generalizability even in complex multi strategy poisoning attack scenarios. \par

\section{Experimental Results}
\label{sec:experiment}

\subsection{Datasets and Experimental Setup}
For this experiment, we conduct our evaluation on a machine equipped with an Intel(R) Xeon(R) W-2133 CPU @ 3.60GHz and 30 GB RAM. The system utilizes an NVIDIA Quadro RTX 5000 GPU. The operating system is Ubuntu. Our experiments are conducted using two widely recognized benchmark datasets: LibriSpeech and VoxCeleb. The LibriSpeech dataset comprises 2,218 user accounts, whereas the VoxCeleb dataset includes 988 user accounts. Additionally, we use the mergerd version of both datasets which comprises 3206 user accounts. We train the CNN model incorporated in our framework on 1,684 users' accounts from LibriSpeech,784 users' accounts from VoxCeleb and 1923 accounts from merged dataset. Consequently, we test the CNN model on 534 users' accounts from LibriSpeech, 204 users' accounts from VoxCeleb and 641 accounts from merged dataset. This experimental design simulates realistic adversarial conditions, allowing us to assess both the PBSM backdoor detection performance and the robustness of our framework under diverse attack scenarios at the same time. \par

\subsection{Evaluation Metrics}
   Firstly, we assess the ability of our framework to correctly classify user accounts based on their acoustic features, emphasizing the separation between triggered and poisoned accounts. Similar to \cite{10538215} ASR is considered as the main metric to evaluate the performance of our PBSM backdoor detection mechanism against the BTA attack.Therefore, we are able to make a relevant comparison in terms of efficiency between our PBSM detection mechanism and the existing ones.
   Furthermore, in order to assess the performance of our CNN model in detecting TDPA, we use classification metrics. We use a separate test dataset for the evaluation of our framework's performance. The trained model is used to predict the class labels for each test file. \par

\subsubsection{User-Level PBSM backdoor detection Performance}
\begin{figure*}[!t]
\centering
\subfloat[]{\includegraphics[width=0.40\textwidth]{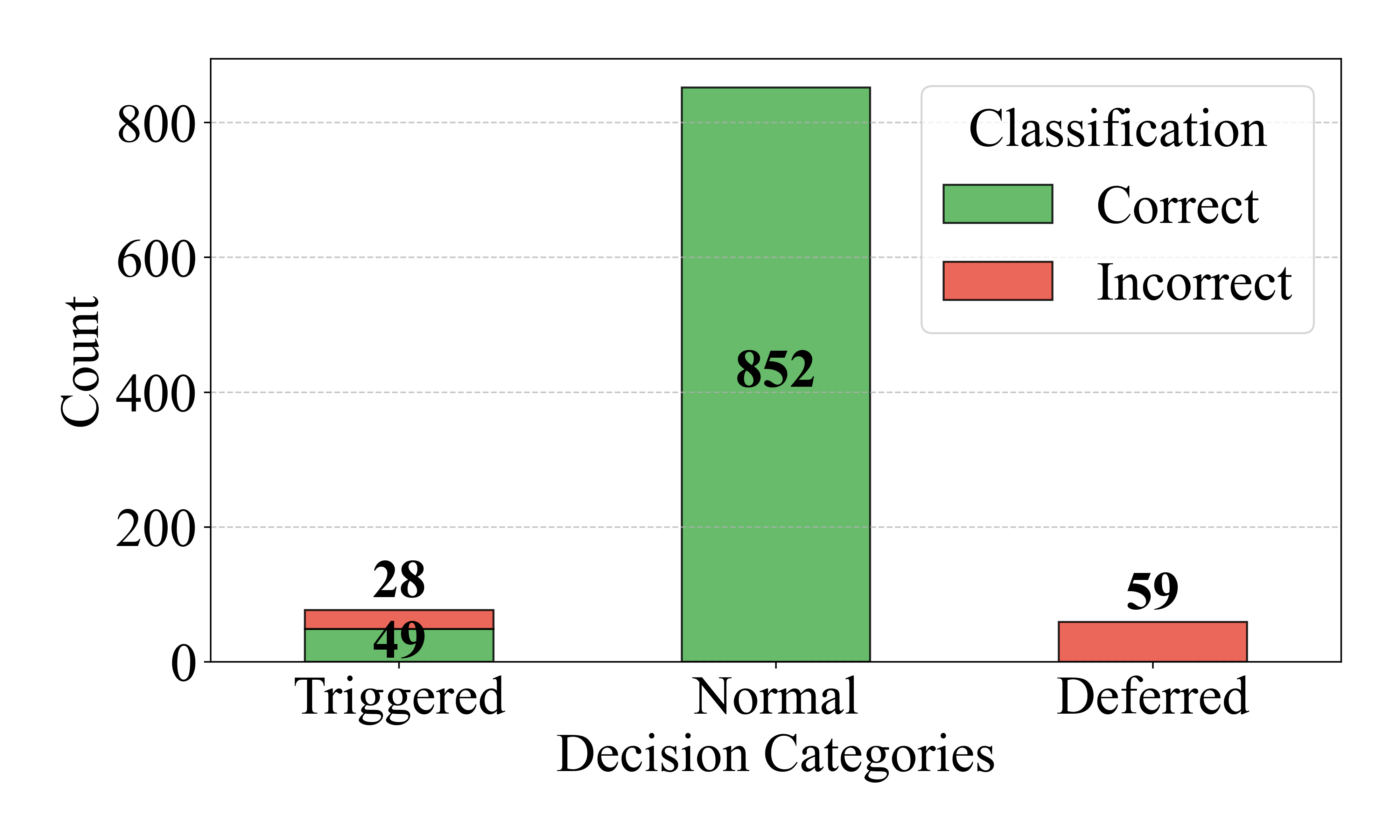}%
\label{fig:voxceleb_stacked_decision}}
\hfil
\subfloat[]{\includegraphics[width=0.40\textwidth]{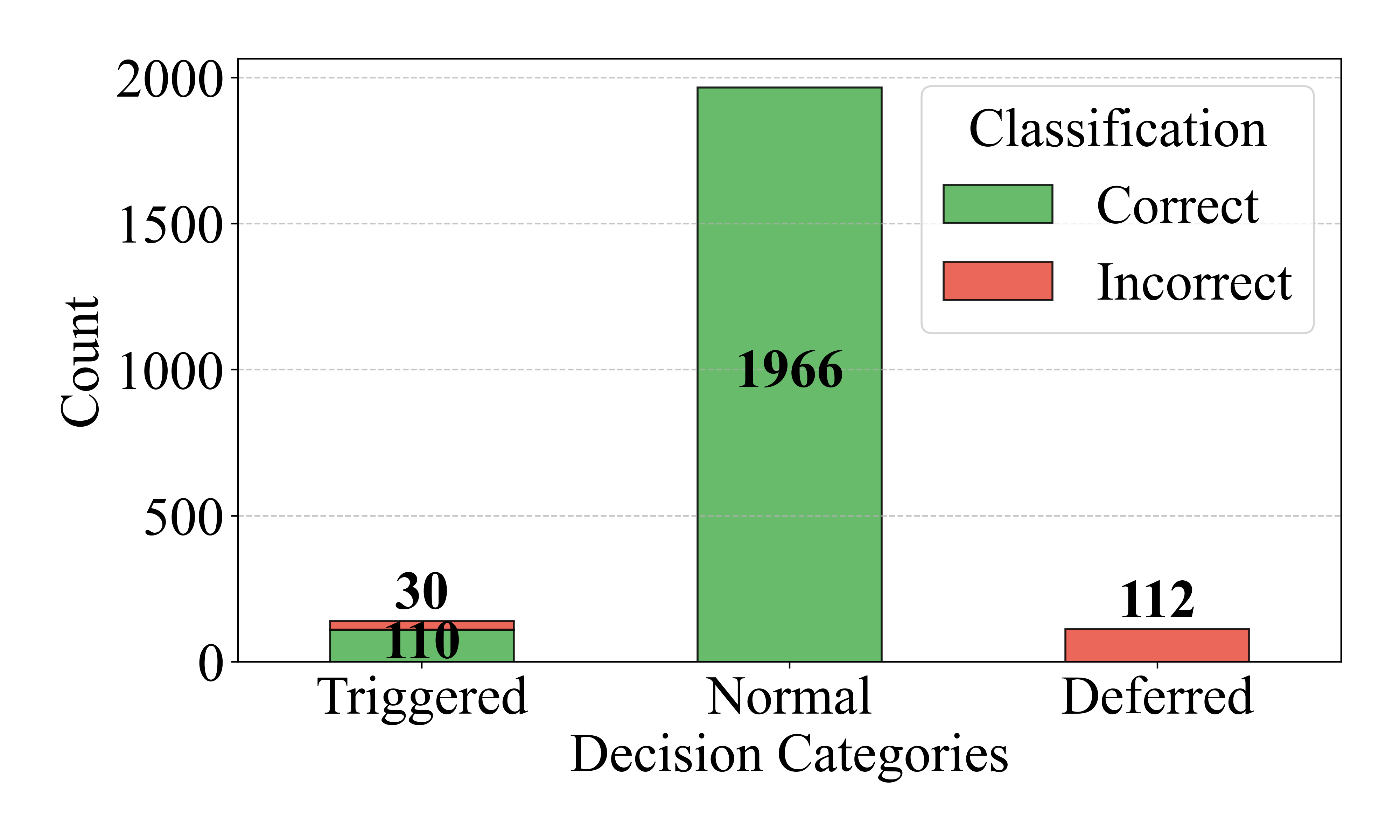}%
\label{fig:librispeech_stacked_decision}}
\caption{Stacked‑bar visualisation of user‑level classification outcomes for the two evaluation. Each bar represents the complete test set for a corpus and is subdivided along the horizontal axis into three decision classes—\emph{Triggered}, \emph{Legitimate}, and \emph{Deferred}.  
    Within each class segment, the green portion indicates accounts whose ground‑truth label matches the automated decision; the red portion marks mismatches.  
    Absolute account counts are printed above each segment; bar height is normalised to the total number of accounts in the respective corpus.}
    \label{fig:stacked_decision}
\end{figure*}

Our framework demonstrates robust performance in classifying user accounts based on their acoustic features. 
Figures~\ref{fig:stacked_decision}(a) and ~\ref{fig:stacked_decision}(b),  present stacked bar charts categorizing user accounts into three decision outcomes: Triggered, Legitimate, and Deferred.

The results directly supports our hypothesis that backdoor modifications—particularly pitch and HF energy shifts—are reliably captured by our detection mechanism, ensuring that no triggered user is overlooked. The legitimate category exhibits strong precision, as indicated by the large green bars. Specifically, 852 legitimate accounts in Figure~\ref{fig:stacked_decision}(a) and 1,966 legitimate accounts in Figure~\ref{fig:stacked_decision}(b) are correctly classified, highlighting the accuracy of our framework. This outcome demonstrates that legitimate users, who generally exhibit moderate pitch and HF-energy levels, are reliably recognized as legitimate. A small fraction of legitimate accounts—28 in Figure~\ref{fig:stacked_decision}(a) and 30 in Figure~\ref{fig:stacked_decision}(b)—appear in the triggered or “incorrect” section. These misclassifications likely result from atypical pitch or HF-energy signatures caused by artifacts such as background noise, unusual vocalization, or partial silence. Additionally, the deferred category—59 accounts in Figure~\ref{fig:stacked_decision}(a) and 112 in Figure~\ref{fig:stacked_decision}(b)—captures cases where acoustic signatures are ambiguous enough to warrant manual inspection. While these deferred accounts predominantly belong to legitimate users, our system conservatively flags them for secondary review, thereby reducing the risk of mistakenly granting access to genuinely triggered users.\par 

\begin{figure*}[!t]
\centering
\subfloat[]{\includegraphics[width=0.38\textwidth]{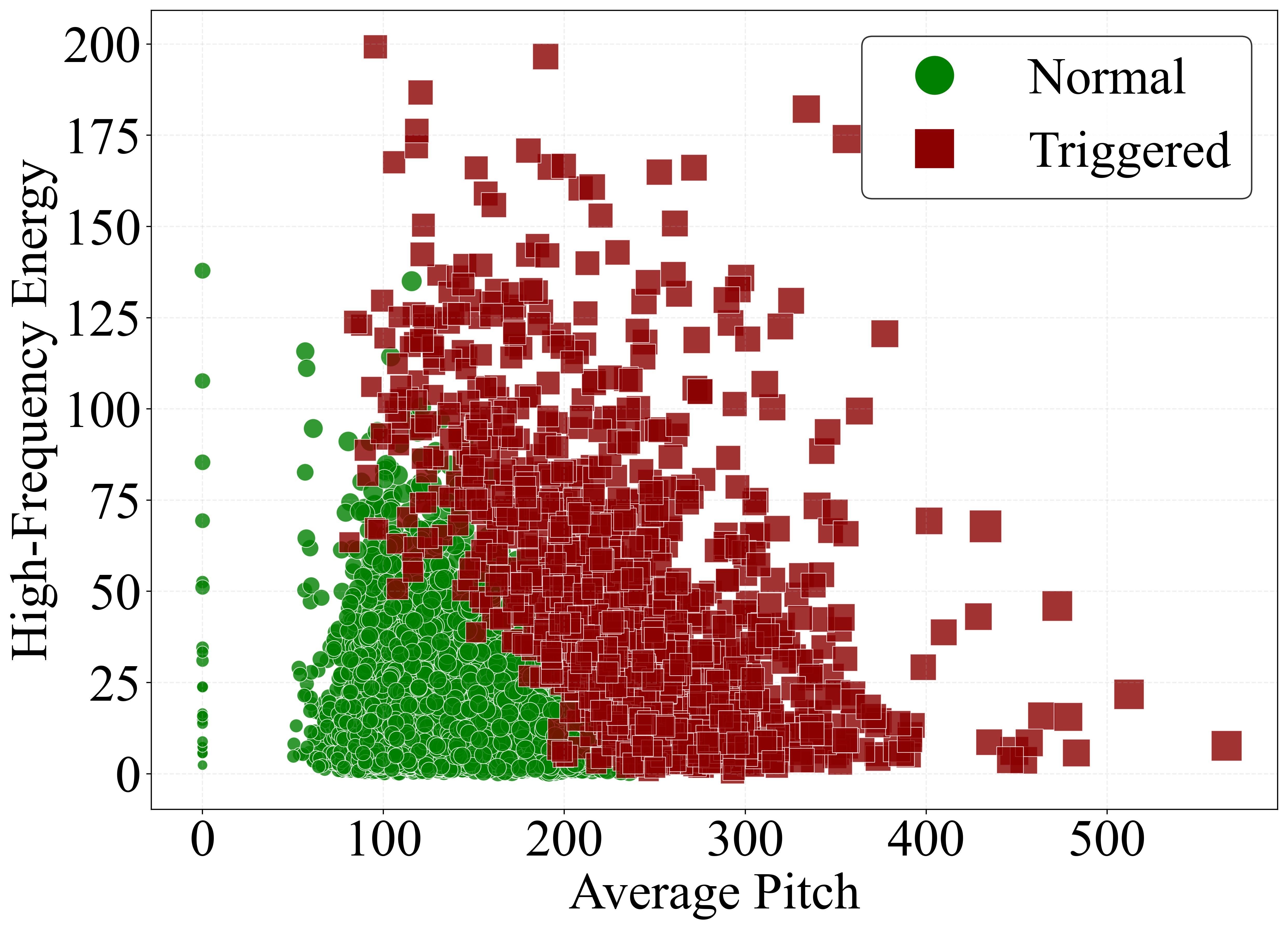}%
\label{fig:voxceleb_scatter_plot}}
\hfil
\subfloat[]{\includegraphics[width=0.38\textwidth]{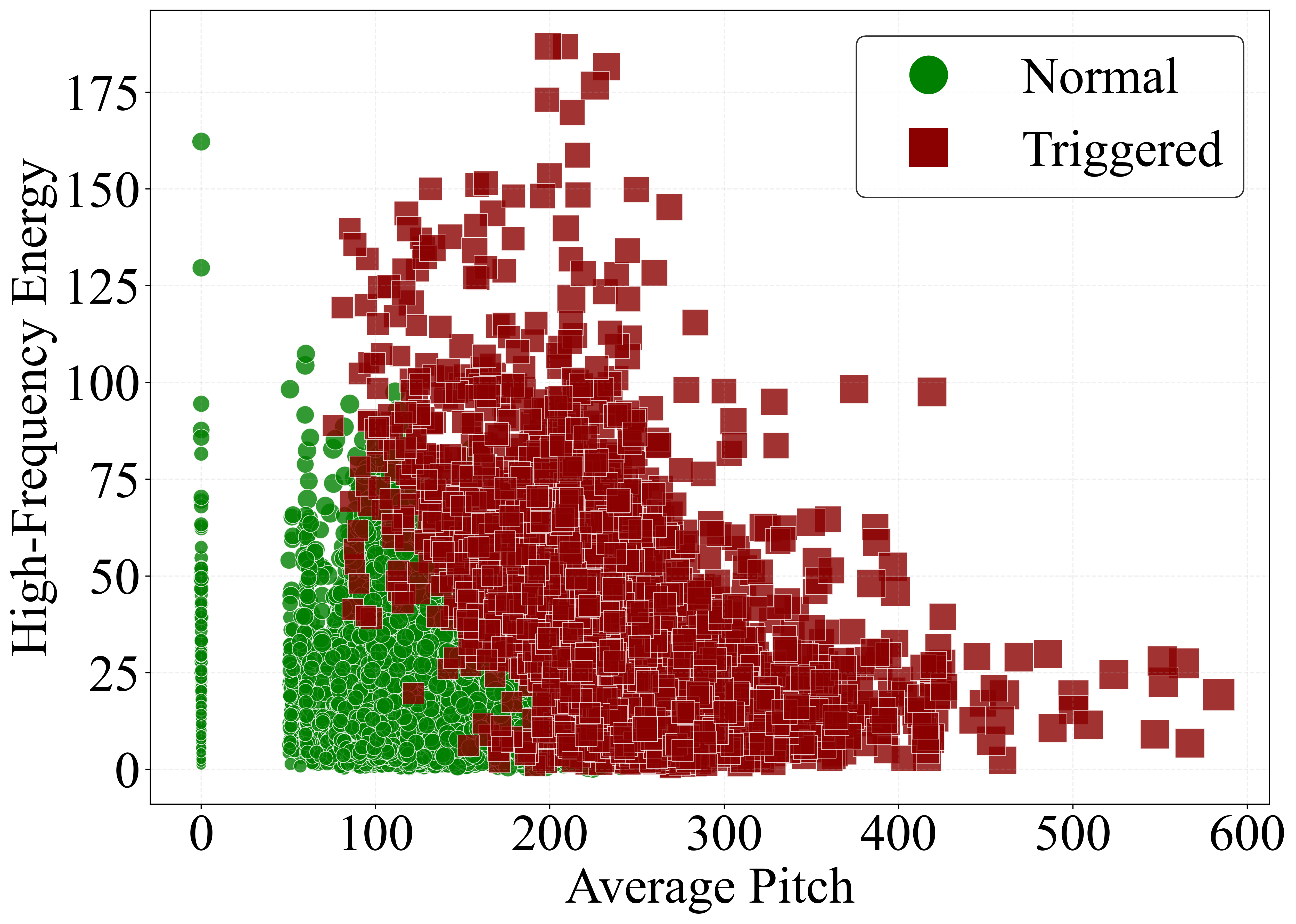}%
\label{fig:librispeech_scatter_plot}}
\caption{
Scatter plots of average pitch versus high‑frequency energy for individual user accounts.  Each point corresponds to a single user account. Green circles represent accounts labeled \emph{Legitimate}; red squares represent accounts labeled \emph{Triggered}.}
\label{fig:scatter_plots}
\end{figure*}

Complementing the classification statistics, the scatter plots are presented in Figures~\ref{fig:scatter_plots}(a) and Figure~\ref{fig:scatter_plots}(b) displaying the average pitch (x-axis) with high-frequency energy (y-axis) for the user accounts. 
A key observation from these scatter plots is the distinct separation in the feature space; as mentioned in the caption of this Figure. 
This empirical evidence supports the robustness of our proposed PBSM backdoor detection in capturing pitch-based manipulations. \par

\subsubsection{Extended Analysis: Classification Metrics, ASR, and Computational Efficiency}

To further provide a holistic picture of the real-world practicality of our framework, we measure the execution times for each major processing stage, ranging from the staging of the attack scenarios to the final classification outputs. Table~\ref{tab:execution_time} reports the total execution times in seconds for the key steps in the process of detection throughout the framework. These measurements are provided for three scenarios. Separate runs on LibriSpeech and VoxCeleb, as well as a run on a merged version of both datasets.
The timings across datasets indicates that our PBSM backdoor detection mechanism scales effectively. Notably, PBSM backdoor detection consistently requires 4--6 seconds per user processing time, even when accounting for minor variability due to environmental or speaker differences. Although, the total execution time increases with dataset size, the overall processing remains within the mentioned time frame per user. \par

\paragraph{Classification Metrics and ASR}
Table~\ref{tab:dataset_metrics} summarizes the classification performance of our framework at different stages. Specifically, the table presents precision, recall and f1-score for the legitimate, attack, and triggered classes across three datasets. The results indicate that, even with the merged dataset containing data from multiple sources, the recall for the attacked and triggered class remain high while precision drops slighty in comparison to LibriSpeech's metrics. Importantly, in the merged dataset, the ASR is elevated at 15.22\% (compared to 11.11\% on VoxCeleb and 4.17\% on Libri0Speech), which we attribute to the increased acoustic and linguistic diversity. Nonetheless, our framework remarkably reduces the risk of BTA from a baseline ASR of 95--100\%, prior to the implementation of our framework as shown in \cite{10538215}, to levels that remain at 4--15\%. It is important to note that the CNN model evaluation stage takes place after PBSM backdoor detection. This implies that even if a triggered account with minimal chance surpasses PBSM detection with nearly 100\% RTA, there remains an approximately 85--96\% probability that it will be detected through the second stage of analysis by the CNN model. \par

\begin{table}[ht]
    \caption{Execution Time for Staging Attack Scenarios and Data Processing on Libri, Vox, and Merged Datasets}
    \label{tab:classification_results}
    \centering
    \renewcommand{\arraystretch}{1.2} 
    \setlength{\tabcolsep}{8pt} 
    \begin{tabularx}{\columnwidth}{|X|c|c|c|}
\hline
        { \textbf{Task Description}} & { \textbf{Libri (sec)}} & { \textbf{Vox (sec)}} & { \textbf{Merged (sec)}} \\ \hline
        Staging Tageted and Backdoor Attacks & 60   & 32   & 87   \\ \hline
        PBSM Detection (Processing All Users)                 & 8977 & 5201 & 14346 \\ \hline
        NPY and Embedding Generation of Audio Files                       & 1557  & 888  & 4095 \\ \hline
        Total Train and Test Time                                    & 3712.38 & 2569 & 5531 \\ \hline
    \end{tabularx}
\label{tab:execution_time}
\end{table}

\begin{table}[ht]
    \centering
    \caption{Precision, Recall, and F1 scores}
    \label{tab:dataset_metrics}
    \renewcommand{\arraystretch}{1.2} 
    \setlength{\tabcolsep}{7pt} 
    \begin{tabularx}{\columnwidth}{|X|c|c|c|c|c|}
        \hline
        { \textbf{Dataset}} & { \textbf{Class}} & { \textbf{Precision}} & { \textbf{Recall}} & { \textbf{F1}} & { \textbf{ASR}} \\ \hline

            & Legitimate    & 0.99 & 0.98 & 0.98 & \\ \cline{2-5}
        {{LibriSpeech}} 
            & Attack    & 0.76 & 0.95 & 0.84 & 4.17\% \\ \cline{2-5}
            & Triggered & 0.81 & 0.98 & 0.88 & \\ \hline
 
            & Legitimate    & 0.96 & 0.95 & 0.95 & \\ \cline{2-5}
        {{VoxCeleb}}
            & Attack    & 0.68 & 0.94 & 0.78 & 11.11\% \\ \cline{2-5}
            & Triggered & 0.71 & 0.96 & 0.81 & \\ \hline

            & Legitimate    & 0.99 & 0.97 & 0.98 & \\ \cline{2-5}
        {{Merged}} 
            & Attack    & 0.72 & 0.93 & 0.84 & 15.22\% \\ \cline{2-5}
            & Triggered & 0.73 & 0.96 & 0.82 & \\ \hline
    \end{tabularx}
    \vspace{-0.2em}
\end{table}

The evaluation of the merged dataset shows that our unified framework significantly reduces the ASR of BTA, while detecting accounts under TDPA with a high recall. With the ability to process each user's audio files in 4-–6 seconds, the PBSM backdoor detection mechanism is very successful in identifying and flagging backdoor triggered audio files at the time of user enrollment. This rapid-response mechanism safeguards the training pipeline from compromised data, preserving high classification recall. Subsequently, our CNN model's training includes triggered users along with attacked ones, which enables it to effectively mitigate backdoor-triggered user accounts and improve overall system resilience, while effectively recognizing TDPA with a high recall. \par

\begin{figure*}[ht]
    \centering
    \subfloat[]{
        \includegraphics[width=0.23\textwidth]{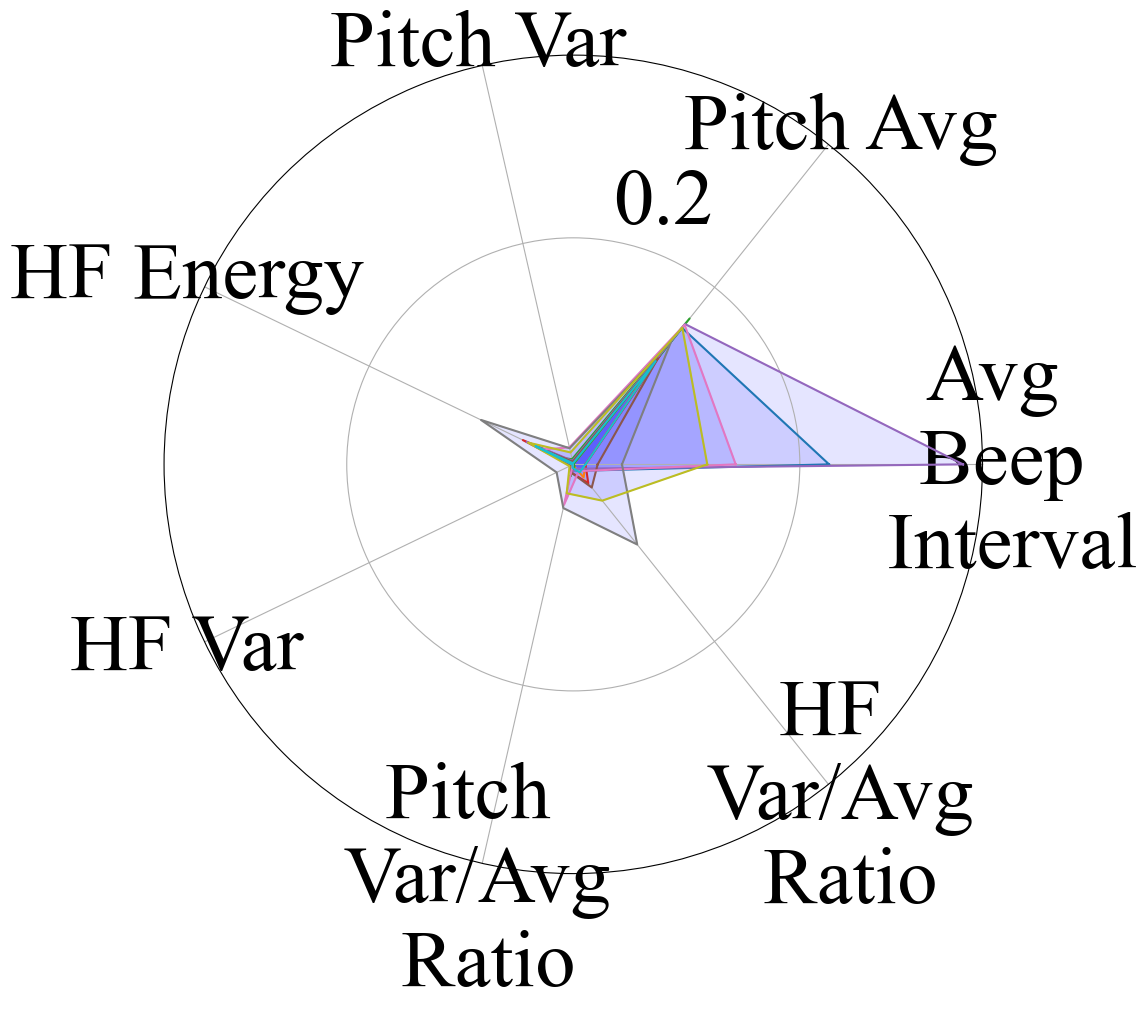}
    }
    \hfill
    \subfloat[]{
        \includegraphics[width=0.23\textwidth]{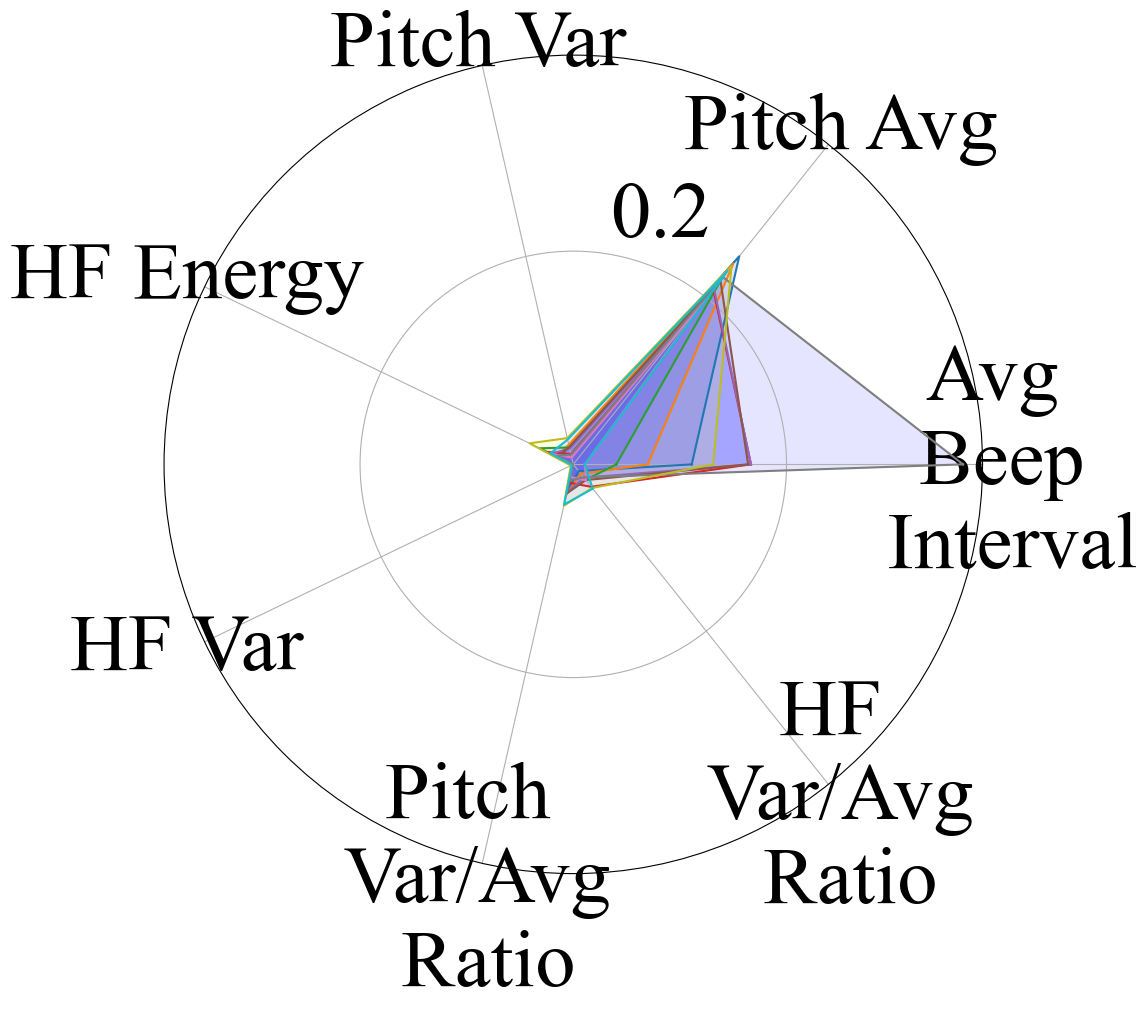}
    }
    \hfill
    \subfloat[]{
        \includegraphics[width=0.23\textwidth]{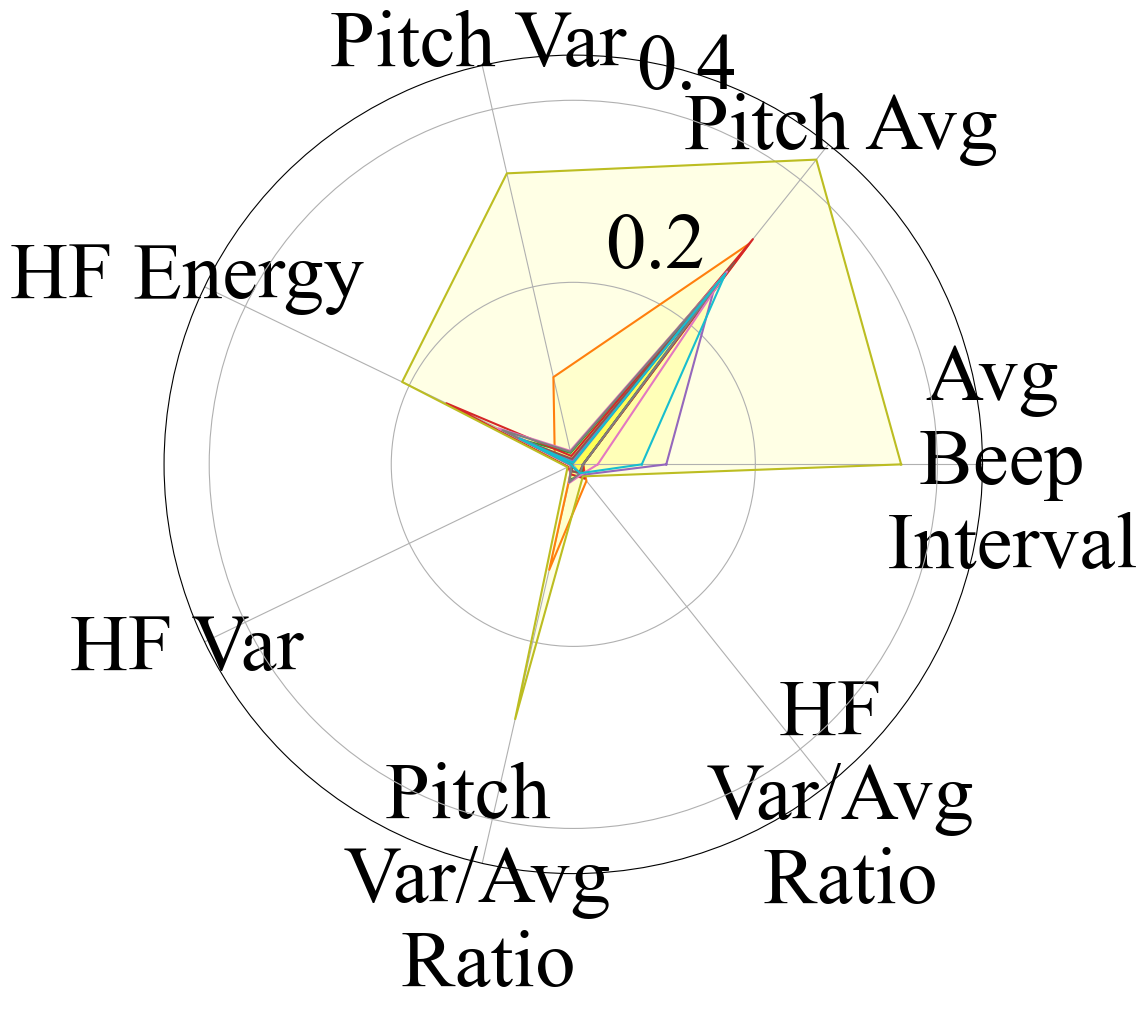}
    }
    \hfill
    \subfloat[]{
        \includegraphics[width=0.23\textwidth]{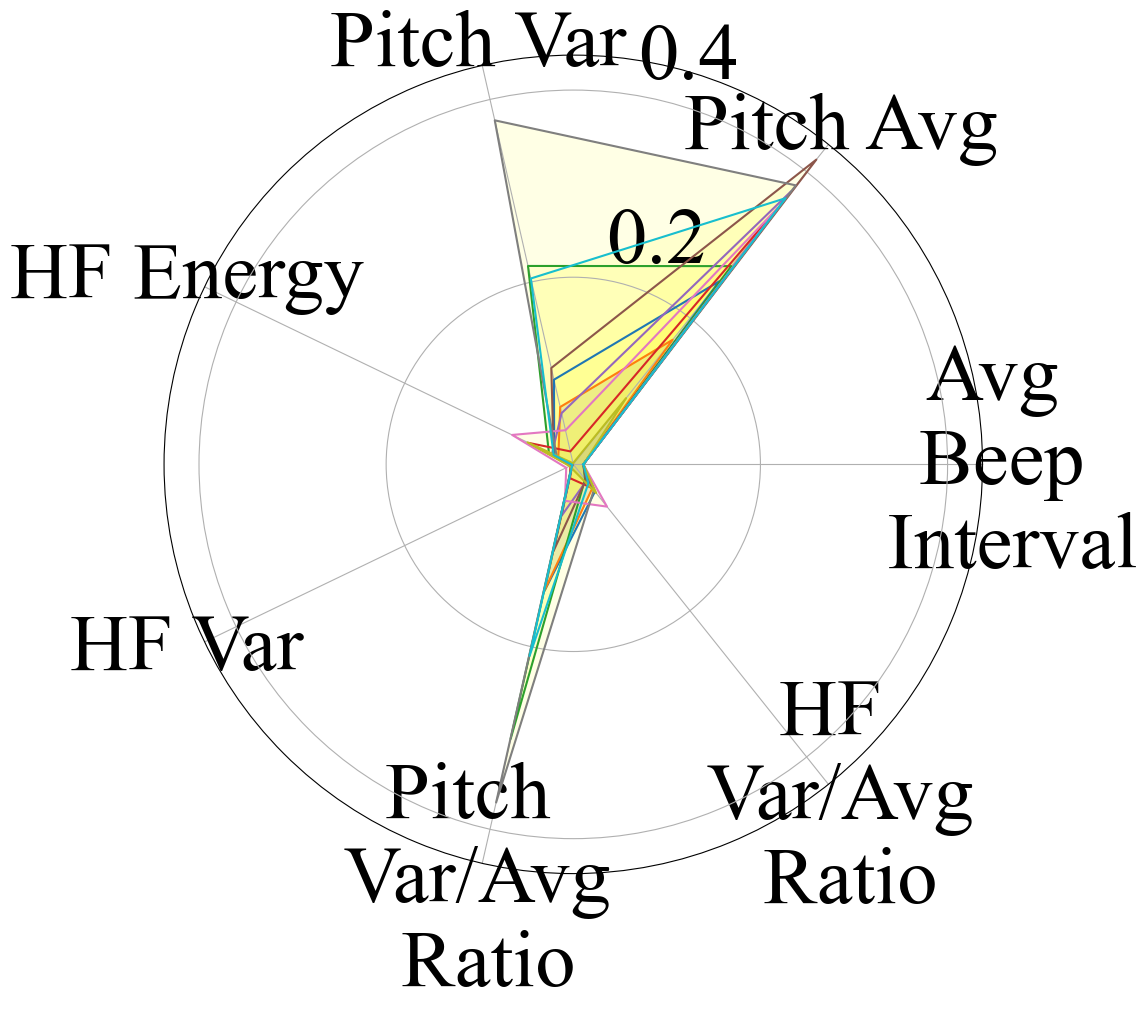}
    }
    \caption{Radar plots of the five normalised acoustic features—average pitch, pitch variance, high‑frequency (HF) energy, HF‑energy variance, and average beep interval—for four representative user accounts.  The dashed grey circle marks the acceptance band \([-1,\,1]\) used in Eq.~(\ref{eq:score}); polygons that extend beyond this band (subfigures(c) and (d)) correspond to accounts whose aggregated score exceeds the trigger threshold~$\tau$.}
    \label{fig:radar_plots}
\end{figure*}

Figure~\ref{fig:radar_plots} presents a radar plot analysis of the multi-dimensional acoustic feature profiles corresponding to both legitimate and triggered user accounts. Each spoke in the radar plot represents a $z$-scored acoustic feature, with the region between $[-1, 1]$ marked as the acceptance band. A user account is classified as \emph{Triggered} if any of its normalized feature means extend beyond this band, as such a deviation ensures that the per-sample score $S(\mathbf{x})$, defined in Eq.~(\ref{eq:score}), surpasses the global threshold~$\tau$.
Subfigures~\ref{fig:radar_plots}a and \ref{fig:radar_plots}b remain entirely within the acceptance region and are thus labeled \emph{Legitimate}. In contrast, Subfigures~\ref{fig:radar_plots}c and \ref{fig:radar_plots}d exhibit clear violations along both the average-pitch and high-frequency energy (HF-energy) axes. These excursions lead to $S(\mathbf{x}) > \tau$ for all samples within those accounts, prompting Algorithm~\ref{alg:defense_design} to flag them as \emph{Triggered}.
For legitimate accounts, the radar plot polygons are compact and show strong alignment across key axes, particularly \texttt{avg\_pitch} and \texttt{avg\_beep\_interval}, indicating stable acoustic behavior. Conversely, the radar plots for triggered accounts display significant outward expansion, especially along the \texttt{avg\_pitch} and \texttt{hf\_energy} dimensions. This increased dispersion is consistent with the presence of HFHPS and other high-frequency anomalies, characteristic of BTA-based attacks.

\begin{table}[ht]
    \caption{Classification Results summarizing each user account’s statistics, including mean score, proportion of triggered files, score variance, and final decision.}
    \label{tab:classification_results_trigg}
    \centering
    \renewcommand{\arraystretch}{1.2} 
    \setlength{\tabcolsep}{8pt} 
    \begin{tabularx}{\columnwidth}{|X|c|c|c|c|}
        \hline
        { \textbf{Account}} & { \textbf{Files}} & { \textbf{Mean Score}} & { \textbf{Triggered \%}} & { \textbf{Decision}} \\ \hline
        
        {0868\_t} & 10 & 123.72   & 100 & \textbf{Triggered} \\ \hline
        
        {1373\_t} & 10 &  125.66 & 100 &  \textbf{Triggered} \\ \hline
        
        {47} & 10 & 90.17 & 11 & \textbf{Legitimate} \\ \hline
        
        {27} & 10  & 63.62 & 0 & \textbf{Legitimate} \\ \hline
        
    \end{tabularx}
\end{table}

Table \ref{tab:classification_results_trigg} depicts the confirmation of our beep-based and weighted-score logic. Although radar charts visually depict anomalies, it is the aggregated proportion\_triggered and mean\_score that unify these observations into a single classification outcome. 

\subsection{Broader Security Implications}

\begin{table*}[htbp]
\centering
\caption{Comparative Summary of Existing Defenses vs. Our Proposed Unified Framework}
\label{tab:ComparativeDefenses}
\renewcommand{\arraystretch}{1.2} 
\setlength{\tabcolsep}{6pt}       
\begin{tabularx}{\textwidth}{|c|c|c|X|X|}
\hline
{ \textbf{Framework}} & { \textbf{Mechanism}} & { \textbf{ ASR}} & { \textbf{Approach}} & { \textbf{Limitations}} \\ \hline

{Fine-Tuning}~\cite{8119189} & BTA Only & $\sim$45\% & Retrains network on a subset of clean data to reduce Trojan activation. & High computational overhead; no real-time capability; lacks frequency-based analysis. \\ \hline

{Model Pruning}~\cite{10.1007/978-3-030-00470-5_13} & BTA Only & $\sim$65\% & Removes dormant neurons to limit backdoor activations. & Degrades benign accuracy; does not analyze pitch/frequency patterns. \\ \hline

{Trigger Filtering}~\cite{10538215} & BTA Only & $\sim$65\% & Amplitude or energy-based filtering. & Ineffective against PBSM; relies on static thresholds. \\ \hline

{TED}~\cite{Mo2023RobustBD} & BTA Only & -- & PCA-based outlier detection across layer activations. & High computational cost; lacks frequency-specific insights; poor scalability. \\ \hline

{Guardian}~\cite{10163863} & TDPA Only & -- & CNN-based discriminator with multi-model training. & High training cost; slow large-scale deployment. \\ \hline

\textbf{Ours} & BTA + TDPA & 5--15\% & Lightweight CNN with frequency-analysis. & Unified detection for BTA and TDPA; real-time processing; preserves accuracy. \\ \hline

\end{tabularx}
\end{table*}

Table~\ref{tab:ComparativeDefenses} presents a comparative analysis of existing defense mechanisms against PBSM backdoor attacks, highlighting their strengths and limitations relative to our unified framework. Unlike prior approaches, our method simultaneously mitigates both BTA and TDPA, achieving lower ASR with minimal computational overhead while ensuring real-time adaptability. This scalability makes our detection mechanism a robust solution for securing voice authentication systems against PBSM-based adversarial threats.
Several mitigation strategies have been explored independently~\cite{8119189, 10.1007/978-3-030-00470-5_13, guo2023scaleupefficientblackboxinputlevel, 10.1109/TIFS.2023.3268882, pmlr-v202-xiang23a}. Some defenses target BTA exclusively, while others focus solely on TDPA. However, none provide comprehensive protection against an adversary employing both attack vectors simultaneously.  \par
Fine-tuning~\cite{8119189} has shown partial success in countering PBSM backdoor attacks by reducing ASR to 45\%, but it demands significant computational resources, requires multiple retraining epochs, and remains impractical for real-time voice authentication. Moreover, it does not explicitly address pitch-based acoustic triggers, limiting its effectiveness against sophisticated poisoning attacks. Model pruning~\cite{10.1007/978-3-030-00470-5_13} weakens backdoor activations by removing dormant neurons. However, this method only reduces ASR to approximately 65\%, leaving systems vulnerable to spectral manipulations such as pitch boosting. Both fine-tuning and pruning fail to integrate frequency- or pitch-aware defenses, a key limitation that our approach overcomes by detecting adversarial triggers before they influence the model. \par
Trigger filtering~\cite{10538215} achieves ASR reductions between 45--65\%, yet it lacks the adaptability required for detecting real-time voice-based attacks. TED~\cite{Mo2023RobustBD}, while effective for image-based backdoor detection, lacks explicit mechanisms for detecting high-frequency or pitch-shifted manipulations central to PBSM attacks. 
Guardian~\cite{10163863} employs multiple neural network models to detect TDPA, significantly increasing training costs and inference latency. In contrast, our framework unifies BTA and TDPA detection through a frequency-based PBSM detection mechanism followed by a lightweight CNN model. \par

The comprehensive experimental evaluation validates the efficacy of our PBSM backdoor detection mechanism in a number of dimensions.  
The efficacy of our proposed CNN-based classification model is further confirmed by ASR and classification metrics assessments. BTA baselines is decreased from 95--100\% before implementing our PBSM backdoor detection to as low as 4.17\% (LibriSpeech), 11.11\% (VoxCeleb), and 15.22\% (merged) after incorporating the detection mechanism. Furthermore the attacked accounts under TDPA have been recognized by our CNN model with a recall as high as 95\% (LibriSpeech), 94\% (VoxCeleb), and 93\% (merged). 
Additionally, timing analysis verifies that every phase of our system functions within realistic execution bounds, which offers its scalability for practical implementation. By identifying separate auditory signatures, the radar plot analysis (discussed in the next section) further supports the differentiation between triggered and legitimate accounts. The differentiation confirms the efficacy of our multi-layer identification mechanism. All together, these results confirm the practical viability and robustness of our framework, which preserves high classification recall across various datasets. \par

\section{Conclusion and Future Work}
\label{sec:conclusion}
This work introduced a unified defense framework capable of detecting covert pitch-boosting backdoor triggered attacks and mitigating data poisoning attack simultaneously. By addressing both BTA and TDPA, we establish a more resilient defense mechanism for modern voice authentication pipelines. Unlike conventional methods that address BTA or TDPA in isolation, our approach integrates an effective detection mechanism against PBSM backdoor attacks by reducing ASR to 5-15\% and a CNN-based classifier that identifies poisoned audio files with more than 95\% recall across multiple datasets. In contrast to existing defenses that impose high computational overhead, our method requires no extensive model re-training or pruning.
Despite these advancements, future research will extend our detection approach to explore adaptive threshold tuning and dynamic feature weighting that can enhance robustness by accommodating heterogeneous user profiles and environmental variations. \par

\bibliographystyle{IEEEtran}
\bibliography{references}

\end{document}